# Efficient Peer-to-Peer Content Sharing for Learning in Virtual Worlds

**Bingqing Shen, Jingzhi Guo**
(University of Macau, Macau, China
daniel.shen@connect.um.edu.mo, jzguo@um.edu.mo)

**Abstract:** Virtual world technologies provide new and immersive space for learning, training, and education. They are enabled by the content creation and content sharing function for allowing users to create and interoperate various learning objects. Unfortunately, virtual world content sharing based on persistent virtual world content storage, to the best of our knowledge, does not exist. In this paper, we address this problem by proposing a content sharing scheme based on Virtual Net, a virtual world persistency framework. For efficient content retrieval, three strategies have been proposed to reduce communication overhead and content load delay. By integrating these strategies, a virtual world content search and retrieval algorithm has been devised. The experiment results verify the effectiveness of the algorithm.

**Categories:** L.3.0, L.6.0, L.6.1, L.6.2, C.2

## 1    Introduction

Virtual worlds have attracted millions of people around the world. The platforms, such as Second Life (secondlife.com), and Sansar (sansar.com), can extend the form of in-classroom learning to interactive learning [Dickey, 2005], experiential learning [Jarmon et al., 2009], immersive learning [De Freitas et al., 2010], social learning [Smith and Berge, 2009], and constructivist learning [Gül et al., 2008] by simulating real-world environments [Kim and Ke, 2016] and offering more forms of collaboration [Burton, 2010] and interaction [Ha and Fang, 2018]. One key enabling element is content sharing. Different from multiplayer online games (MMOGs), users can utilize their expertise and creativity to create discoverable, reusable, and sharable learning objects [Wiley, 2002]. Through augmented reality and internet of things, for example, context-aware learning objects can be created and distributed at different locations for student discovery with mobile applications [Sampson et al., 2013]. Moreover, wearable virtual reality allows museum visitors to create and share personalized content for collaborative touring [Kosmopoulos and Styliaras, 2018].

To support the core characteristic of reusability [Wiley, 2002], learning objects should be persistently maintained. Thus, for content sharing, a virtual world, together with all the generated contents, should exist forever regardless of any change on virtual world owners or users. Currently, however, nearly all existing virtual worlds are created and owned by certain entities (often commercial companies), and user-generated contents are stored on the servers of these entities. When the entity owning a virtual world dies, bankrupts, or withdraws its operations, the affected virtual world will collapse[1] together with the loss of the user-created contents.

To protect virtual world users and their created contents, the persistency feature of virtual world must be maintained. In the previous research [Shen et al., 2017], a peer-to-peer (P2P) computing [Schollmeier, 2001] platform, called Virtual Net, has been introduced to prevent the possible collapse of virtual world and maintain the persistency feature. The central idea of Virtual Net is that nobody owns the virtual world but all users collectively create a self-organized virtual world, so that it will not fail due to the departure of any entity. In this approach, each virtual world user contributes a part of his/her computing resources, including a certain amount of CPU time, memory, storage, and bandwidth of his/her computing device(s). Devices can be smart phones, personal computers, etc. The contributed part of each device is then virtualized into one or multiple nodes. Each user of Virtual Net is then assigned one or multiple nodes which can store his/her contents or deploy some applications without a central management.

Content sharing is challenging without the coordination of a central server. It can be divided into two steps: content storage and content retrieval. The first step, reliable content storage over Virtual Net, has been studied in [Shen, 2017]. The second step will be discussed in this paper. In virtual world content sharing, content discovery is location-based. Users can dynamically perceive new contents along with their movement. Also, content retrieval needs to be complete. When multiple users access to a shared environment, they can interact with the same set of objects in the environment.

Virtual world content retrieval has been studied in [Symborski, 2008] and [Symborski, 2010] for reducing the communication cost of servers in content distribution. These works utilize summary descriptor to achieve content discovery completeness. The design is based on the central-server model. On the other hand, P2P file sharing [Lua et al., 2005] has been intensively studied within the last decade. But most of them do not support location-based query. P2P semantic file search [Gupta et. al, 2003], [Tang et al., 2003] allows file retrieval based on range query by providing multiple semantic variables, which is most related to location-based content retrieval. However, semantic file search do not require search completeness. Thus the solutions cannot be applied in virtual world content retrieval.

In this paper, we provide a new content sharing scheme tailored for virtual world over Virtual Net. The basic idea is to provide each user a list of contents in a virtual world. A user client then downloads the contents from the nodes of Virtual Net. A new service, called object resource lookup, is devised for looking up the network address of nodes. The basic scheme is, however, inefficient, since there could be many contents to be download, leading to three challenges: 1) high content load delay, 2) redundant content retrieval, and 3) high communication overhead.

We then address these problems by proposing three strategies. The proximity-based content retrieval strategy searches and loads nearby contents first, followed by pre-loading the remote contents within a certain range. The region-based content inventory is designed for local cache management, which can eliminate redundant object retrieval. Lastly, the location mapping service is designed to reduce communication overhead by caching the actual content storage location. The experiment results have validated the effectiveness of these strategies.

The remainder of the paper is organized as follows. Section 2 provides the preliminary techniques used in the solution design, including a brief introduction of the Virtual Net Architecture. Section 3 gives an overview of virtual world content sharing and the related problems. Section 4 solves the problems by providing the efficiency improvement strategies. Section 5 provides the algorithm design to integrate and implement the strategies. Section 6 evaluates the algorithm with experiments. Related work is discussed in Section 7. Lastly, Section 8 concludes the paper.

---

[1] The inactive virtual world list in http://opensimulator.org/wiki/Grid_List.

## 2 Preliminaries

In this section, some preliminary techniques are introduced for supporting the design of the content sharing scheme. First, since our scheme is devised based on Virtual Net, its architecture is briefly described. Then, two P2P overlay networks, CAN and Chord, are introduced, which are utilized as the building blocks in our solution. They can provide the function of distributed hash table (DHT) [Lua et al., 2005] to reliably store information without a central server.

### 2.1 Virtual Net Architecture

In the Virtual Net architecture [Shen et al., 2017], the user contributed devices are virtualized into one or more equally capable virtual machines, called nodes. That is, all nodes have the same amount of computing, storage, and communication capacity. Each user is assigned one or several nodes for storing his/her virtual world applications. All the nodes of a user forms a computing unit, called logical computer.

To overcome node failure, the contents on each node in a logical computer is replicated to multiple nodes, called replica nodes or replicas. Multiple replica nodes hosting the same contents form a replica group. In a replica group, if the number of surviving replicas is less required, new replicas will be created by the surviving replicas. Thus, a replica group can survive by monitoring the state of the replicas and repeating the replication cycle.

The logical computer model can not only support reliable content storage, but also support reliable program execution [Shen and Guo, 2018]. In a virtual world application, a user can interact with his/her virtual objects by sending events to all the replicas. Through a fast consensus protocol [Shen and Guo, 2018], all the replicas can deliver the events in the same sequence so that they always maintain the same state. Thus, any failure of a node will not interrupt the user playing in the virtual world.

### 2.2 Content-Addressable Network (CAN)

CAN [Ratnasamy et al., 2001] is a P2P overlay network protocol. As illustrated in Figure 1 (a), the geometry of a CAN overlay is a d-dimensional Cartesian space, with each dimension normalized to 1. The entire space is dynamically partitioned into zones and each zone is assigned to a peer node. Each object is assigned a location with a deterministic function converting the object ID to a *n*-d coordinate. The object is stored on the node whose zone range covers the object coordinate. Thus, locating an object is reduced to routing to the node storing the object. Routing in a CAN overlay involves traversing from one zone to another. Thus, the routing length is $O(n^{1/d})$ where *n* is the size of the overlay. When a new node joins the overlay network, it will randomly picks a coordinate. Then, the node will be routed to the zone covering the coordinate, and equally split the zone with the current owner.

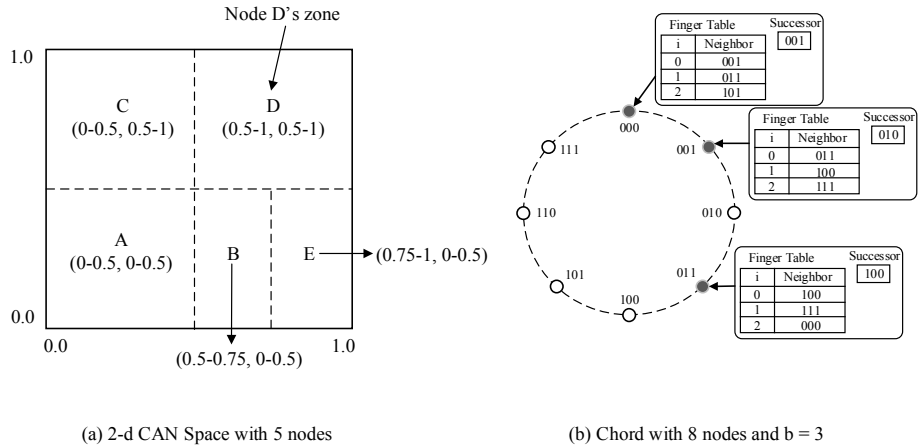

(a) 2-d CAN Space with 5 nodes          (b) Chord with 8 nodes and b = 3

*Figure 1: Examples of (1) CAN overlay and (2) Chord overlay*

### 2.3 Chord

Chord [Stoica et al., 2001] is another P2P overlay network protocol. Each node is assigned a unique hash key of *b* bits as the node ID. The ID space is partitioned into multiple segments. One node is the predecessor of another if it has the largest node ID smaller than the successor among all nodes. Then, the predecessor is assigned the segment from its ID to the ID of its successor. Moreover, the node with the smallest ID is the successor of the node with the largest ID. Thus, the basic network structure of a Chord is a ring, as illustrated in Figure 1 (b). Each node also maintains a finger table which contains *b* entries. For the *i*-th entries, the node maintains some neighbors whose IDs are within the range $[2^i, 2^{i+1})$. An object is also assigned an ID with the same length. It is stored on the node whose segment contains the object ID. Thus, similar to CAN, locating an object is converted to find the node managing the segment. Routing in a Chord overlay can utilize finger tables. Thus, the routing length is $O(\log(n))$ where *n* is the size of the overlay.

### 2.4 Notations

The data formats are defined with basic set and logic notations to facilitate description. Throughout the paper, the symbols for representing the relations of elements are listed in Table 1.

| Notation | Description |
|---|---|
| := | Definition |
| ⟨⋯⟩ | Tuple, representing data or message format |
| {⋯} | Set, representing one or more items of the same type |
| ‖ | String concatenation |
| . | Membership |

*Table 1: Notations*

## 3 Overview of Content Sharing

Different from MMOGs, the contents in a virtual world is mainly created by users [Zhou et al., 2018]. Virtual world contents include various virtual objects, such as user avatar and virtual building. They are presented with 2D or 3D models simulating real- or imaginary-world objects. For rendering, a virtual object (or object in short) is composed of different types of resource files. The basic resource files include the image files and modelling files describing the geometry, texture, and material of the object. For more vivid presentation, some objects also contain animation files, sound files, and script files. Each object can be uniquely identified by object identity and described by some properties, including the name, author, version, etc.

One important property is the object location in a virtual world. It can be described with a 2- or 3-dimensional coordinate. User-created contents are spatially distributed at different places in a virtual world. After a user creates an object, other users can see or interact with it by retrieving the resource files.

A user needs to see the objects within his/her visual range, when he/she enters a new scene. Moreover, when the user moves, new objects need to be discovered. Thus, a mapping between geographic location and object need to be maintained for location-based object query. As illustrated in Figure 2, the mappings of all objects can be maintained by a content list. The network location where the resource files of an object can be downloaded is maintained as the storage address property of the object.

### 3.1 Object Resource Lookup Service

In Virtual Net, objects are stored on the logical computers of their authors. Thus, the storage address property maintains the ID of the object author's logical computer. When a user needs to load an object to a scene, the user client firstly finds the network location of the logical computer by searching against the logical computer ID.

In Virtual Net, some reliable nodes are employed to provide common services which may be needed by all users. These nodes are called soft bots. One type of soft bots provides the object resource lookup service, called addressing bots. They maintain the mapping from a logical computer ID to the addresses of replica nodes. Addressing bots are organized into a Chord overlay, as illustrated in Figure 2. Each addressing bot maintains some hash keys generated from logical computer IDs. The service supports three operations.
1. **Mapping creation**. When a new logical computer is created, a new mapping record is added into the Chord overlay.
2. **Mapping Update**. When a logical computer changes its replica nodes, the mapping record will be updated with the network addresses of the new replica nodes.
3. **Mapping Query**. Given a logical computer ID, the network addresses of the replica nodes are replied to the requester.

The common function shared by the three operations is mapping record lookup. For a given logical computer ID, firstly, a hash key is generated by hashing the ID. With the hash key, the addressing bot maintaining the mapping can be deterministically found through the routing over the Chord overlay with $O(log(n))$ hops on average, $n$ denoting the number of addressing bots.

Through the addressing bots, the replica nodes of a logical computer can be found. Then, the object resource files can be retrieved by requesting it from one of the replica nodes with the object identity.

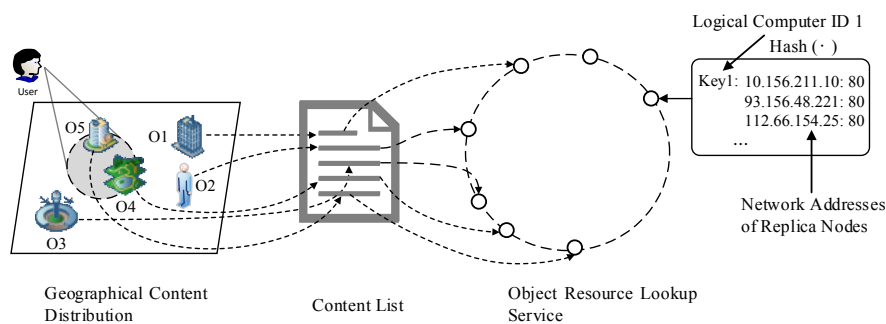

*Figure 2: Illustration of content sharing: Mapping from object location to object storage through the content list and the object resource lookup service*

### 3.2 Content Retrieval

Figure 3 shows the flowchart of the content retrieve process. When a user enters a new scene, there are a bunch of objects distributed at different locations of the simulated space. Firstly, the user client retrieves a complete list of these objects (maybe from a specific logical computer or a server). For each object on the list, the client uses the logical computer ID in the storage address property to find the network address of the replica nodes through the mapping query operation of the object resource lookup service. With the returned set of replica node network addresses, the client then picks one and sends a request to the node for the object resource files. After receiving the object resources, the client renders the object to the GUI. The same process will repeat until all the object on the content list are loaded.

In a virtual world, since contents are created or modified by users from time to time. The above content retrieval process is thus periodically executed, by retrieving the latest content list, to ensure that the objects interacted by users are up-to-date for state consistency among all clients.

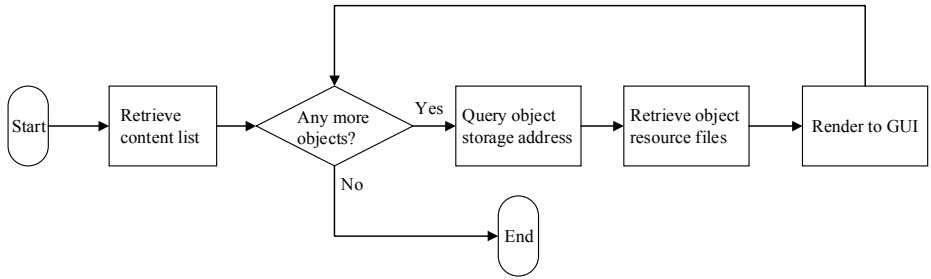

*Figure 3: Flow of basic content retrieval*

### 3.3 Major Challenges

The basic idea of content sharing is straightforward, but there are three major challenges to overcome before it can work efficiently.

**High content load delay**. Following the sequence for object load, some of them remote to the user could be loaded first, while other nearby objects may be loaded later. When the number of objects is large, the user may get aware of the delay of nearby object load. As illustrated in Figure 2, the objects (O4 and O5) within the perception range could be lastly loaded.

**Redundant object retrieval**. Some objects may have already been downloaded before. If they are not modified, they do not need to be downloaded again. On the other hand, if an object has been modified since the last access or some resource files have been corrupted, only the related files need to be retrieved.

**High communication overhead**. Suppose there are $k$ objects to be downloaded. On average, each object retrieval requires $O(log(n))$ hops to find the addressing bot for logical computer lookup, communicating with $O(log(n))$ intermediate bots. Then, searching all the $k$ objects will generate $O(k \cdot log(n))$ communication overhead in total.

These challenges can be concluded that the basic content retrieval idea is inefficient in virtual world content sharing. They will be addressed in Section 4 in detail.

## 4 Efficient Content Retrieval Strategies

To overcome the efficiency problem in content retrieval, objects should be retrieved in the order of their distance to a user, objects should be locally cached and verified to avoid redundant downloads, and the lookup of addressing bots should be minimized to reduce communication overhead. This section introduces the design of these strategies in a consistent way.

### 4.1    Proximity-based Content Retrieval

To reduce the content load delay, the search of the share content is proximity-based in two aspects. First, the contents close to the user will be retrieved and loaded first, then the remote ones. Second, some remote contents within a specific range will be searched after nearby contents for pre-load to minimize the chance of content load delay that a user may be aware of in movement.

Without loss of generality, suppose the content of a virtual world is distributed on a 2D map. Let the lower-left corner of the map be the origin point. The map has the range of [0, X] and [0, Y] respectively along X-axis and Y-axis. The map is partitioned into multiple regions, and each partition is divided into *m* grids. The location of a grid is determined by the coordinate of the lower-left corner of the grid. So is the region coordinate. It is recommended that the size of a grid is no smaller than the user perception range such that a user can retrieve all the perceived contents by loading the contents of at most four grids. Figure 4 illustrates a partition of a 2D map partitioned by 20 regions and 9 grids in each region.

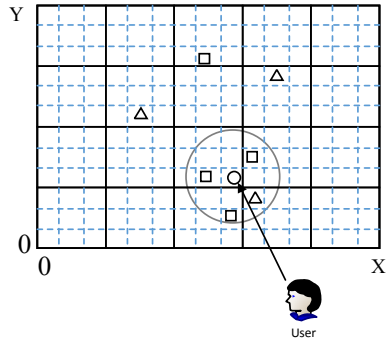

*Figure 4: Map partition of virtual world*

When a user logs into the virtual world, the objects within the grid of user location will be retrieved first, then all the contents located in the neighbour grids. From each neighbour gird, new search routes will be generated to traverse all the unvisited neighbours. These cascaded search routes traverse the map from closing grids to remote grids, allowing nearby contents to be loaded first. Figure 5 illustrates part of the search routes originated from Grid 1.

The proximity-based content search needs to be terminated when a specific range of grids have been traversed, so that a new search can be started with a new origin location. Let $d_r$ be the radius of the pre-defined search range. By appropriately design the region size, a search can terminate when it has already cover the range of a region. For simplicity, suppose the shape of a grid and a region is a square. Let *l* be the side length of a grid. The length of the half diagonal of a region is $d_r = \frac{\sqrt{2}}{2} l \cdot \sqrt{m}$. Let $p_0 = (x_0, y_0)$ be the search origin and $p = (x, y)$ the coordinate of a grid. To cover the range of a region, a search route stops if $|p - p_0| = \sqrt{(x - x_0)^2 + (y - y_0)^2} > d_r$. Figure 4 illustrates the search over the area filled with diagonal lines.

With new content creation and user movement, the above content retrieval process needs to be repeated with cycle length *T*. In each cycle, the central location $p_0$ is changed to the current user location. *T* can be determined by the user movement velocity *v* and the size of a region for minimizing content load delay, e.g., $T = \frac{\sqrt{2}d_r}{v}$.

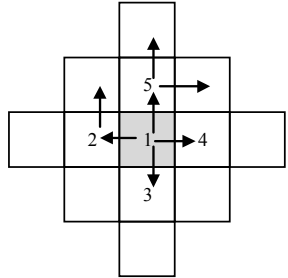

*Figure 5: Example of search routes*

### 4.2 Region-based Content Inventory

Retrieved content can be cached in the local storage to minimize download overhead. If a content has already been retrieved and unchanged since the last access, it does not need to be downloaded again from the network. Thus, only the newly added or modified content resources need to be retrieved. For this purpose, content inventory is needed to maintain content retrieval completeness, content integrity, and content freshness.

The entire content inventory is divided by regions. The devised content inventory scheme hierarchically contains three types of entries: inventory, objects, and files.

**Inventory** Each region has an inventory file, identified by a unique Region ID (*RID*). It contains region coordinate and a set of objects, as in Def. (1). A sample inventory file is illustrated in Figure 6 in JSON.

$$Inventory \coloneqq \langle RID, \{Object\}, RCoord \rangle$$
$$RCoord \coloneqq \langle x, y \rangle \qquad (1)$$

```
{
  "RID": "XLGkgq61BTaQ8NhkcqyU7rLcnSa7dS",
  "RCoord": "900, 800",
  "Object": [<Object 1>, <Object 2>, <Object 3>]
}
```

*Figure 6: Sample inventory*

**Object** An object is identified by the object ID (*OID*), located by its coordinate (*OCoord*) and the logical computer ID (*LCID*), and composed of one or multiple components, as in Def. (2). One indispensable component is *OProperties* which contains the properties of an object, including object name, author, version, etc. Other components are the files related to the object, which are classified into animation, sound, texture, script, etc. Each set of files belonging the same class forms a file type (*FileType*) which includes the type name (*Type*) and the hash code (*FTHash*). *FTHash* is calculated by the Merkle root [Merkle, 1987] of the file hashes (*FHash*) within the *FileType*. The hash code of *Properties* and the hash codes of all *FileTypes* form a Merkle tree [Merkle, 1987] which calculates the object hash code (*OHash*). A sample object description is illustrated in Figure 7 in JSON.

$$
\begin{aligned}
&Object \coloneqq \langle OID, OHash, , OCoord, LCID, OProperties, \{FileType\}\rangle \\
&OProperties \coloneqq \langle Name, Author, Version, \cdots, PHash \rangle \\
&FileType \coloneqq \{\langle Type, FTHash, \{File\}\rangle\} \\
&FTHash \coloneqq MerkleRoot(\{File.FHash\}) \\
&OHash \coloneqq MerkleRoot(PHash, \{FileType.FTHash\})
\end{aligned}
\quad (2)
$$

```
{
    "OID": "2039THflwkjgp0q3pwojfolga",
    "OHash": <Object1's Hash>,
    "OCoord": "101, 203",
    "LCID": "Logical-Computer-100",
    "OProperties:" {
        "PHash": <Properties Hash>,
        "Name": "Object 1",
        "Author": "John Lee",
        ...},
    [{
        "Type": "Texture",
        "FTHash": < Textures Hash>,
        "Files": [ <File 1>, <File 2>]},
    ...]
}
```

*Figure 7: Sample object description*

**File** A file entry contains file hash code (*FHash*) and file properties (*FProperteis*), as in Def. (3). *FProperties* contains the descriptions of a file, including file name, author, version, etc. *FHash* is calculated by hashing the content of a file and its properties.

$$
\begin{aligned}
&File \coloneqq \langle FHash, FProperties \rangle \\
&FProperties \coloneqq \langle Name, Author, Version, \cdots \rangle \\
&FHash \coloneqq Hash(FileContent \parallel FProperties)
\end{aligned}
\quad (3)
$$

Based on the structure of the content inventory, Merkle trees can be hierarchically constructed with the file hash code (*FHash*), component hash code (*PHash* & *FTHash*), and object hash code (*OHash*), which is illustrated in Figure 8. Based on Def. (2), a component hash code (*C*) is constructed by all the file hash codes (*F*) of the component; an object hash code (*O*), which is the root hash of a Merkle tree, is constructed by all the component hash codes of the object. The *P* node represents the *PHash* value, a special component node. With the Merkle tree, the integrity of a parent node is determined by the integrity of its children nodes. With the tree structure, the number of node visit in searching a corrupted node is $O(log_b N)$ where $N$ is the number of files and *b* is the average number of children of a parent node.

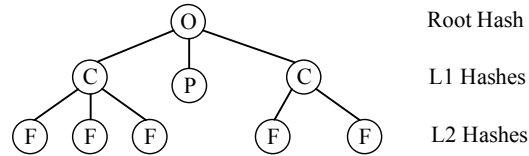

*Figure 8: Content inventory Merkle Tree*

For object modification, the content inventory scheme facilitates object retrieval completeness and content integrity check. All object changes will be reflected on the object files with the updated file hashes which will be aggregated up to the object hash. The content inventories are also maintained in the local storage of clients. After retrieving the content inventory of a region, a user client checks the content integrity by comparing it with the local copy. For each object, if the user finds that the object hash code remains unchanged, no file of the object needs to be retrieved. Otherwise, the object resources will only be retrieved by identifying the modified files through *PHash* comparison, *FTHash* comparison, and finally down to *FHash* comparison.

### 4.3 Location Mapping Service

To reduce communication overhead, in Virtual Net, another type of soft bots, called region bots, is introduced to provide the location mapping service. Each region bot manages one or multiple grids and maintains the inventory of the objects on the grid. For each object, a region bot also maintains the logical computer ID of the author and caches the addressing bot which maintains the network location of the logical computer. Periodically, each region bot looks up the addressing bot for each object against the logical computer ID for cache freshness check. If the addressing bot for maintaining an object has been changed due to bot dynamics, the region bot will update the cache with the new addressing bot.

When a client queries a region bot for the storage address of an object, the bot replies both the logical computer ID and the addressing bot to the client. The client firstly directly queries the object resource from the addressing bot. In case of wrong addressing bot, the client then looks up the correct one against the logical computer ID through the object resource lookup service.

To support proximity-based content storage, all region bots are connected to a 2-dimensional CAN overlay network. The coordinate space of a map is also mapped to the 2-d space of CAN with the following two adaptations. First, a point $p = (x, y)$ on the map is normalized to a point $p'$ in the 2-d space of CAN by $p' = (x / X, y / Y)$ where *X* and *Y* are the side lengths of the map. Second, a space partition will be rounded up to a grid. Figure 9 shows an example of CAN space partition with 5 region bots. For simplicity, units are represented by the number of grids, denoted by *g*. In the example, when Bot 5 joins the zone belonging to Bot 4, Bot 4 horizontally split the zone to 5*g* and 4*g* width, and leave the right side to Bot 5. These adaptations ensure that each region bot manages one or multiple grids.

A 2-d CAN overlay supports search against a coordinate to the region bot in charge of the coordinate with $O(n^{1/2})$ hops on average, where *n* denotes the number of region bots. With the CAN overlay, the location mapping service supports the following five operations.

**Object Creation**. When an object is created, the author's logical computer ID and the addressing bot in charge of the actual storage location will be stored on the region bot in charge of the object coordinate, illustrated in Figure 9 (Object 1).

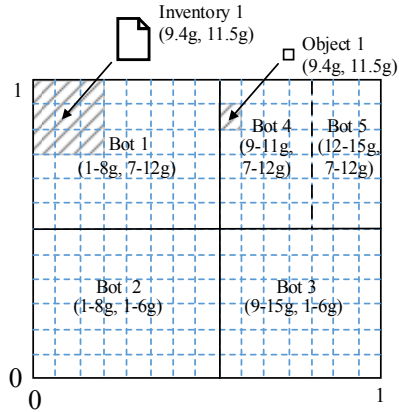

*Figure 9: Example of 2-d CAN space with 5 region bots*

**Content Inventory Creation**. When an inventory is created, it will be stored on the region bot in charge of the region coordinate, illustrated in Figure 9 (Inventory 1).

**Object location Retrieval**. Given the coordinate of an object, the CAN overlay can find the region bot in charge of the object storage information.

**Inventory Retrieval**. Given the coordinate of a region, the CAN overlay can find the region bot in charge of the content inventory of the region.

**Neighbor Query**. A region bot replies up to four bots to the requester, which are the neighbors of the requested bot in the CAN space.

The last operation directly supports the proximity-based content retrieval, as shown in the algorithm design in Section 5.

### 4.4    Discussion

The efficiency of content retrieval can be largely improved with the proposed strategies. First, the proximity-based content retrieval strategy can address the first problem. The pattern of the search routes, like a ripple, can maximize the chance that the objects close to a user will be found first, then the remote ones. By pre-loading some of the remote objects, the strategy can minimize the chance of content load delay.

Proximity-based content retrieval also provides flexibility in cache management. Content cache can be prioritized based on object distance to users. Nearby objects have higher priority than remote objects. If the cache is full, then the low priority objects outside the perception range can be recycled. This is important for resource-limit (e.g., handheld) devices. If their storage space or transmission rate is limited, the perception range will be reduced and the objects within the range will still be firstly retrieved.

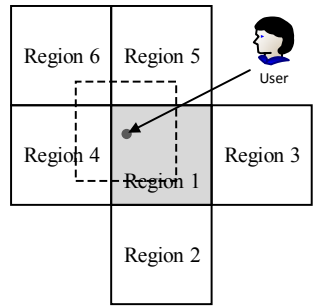

*Figure 10: Example of neighbor region determination*

Second, the content inventory design can reduce the chance of object resource download. By comparing the retrieved content inventory with the local one, the up-to-date objects or object resource files that are already in the local cache can be removed from the download list.

Third, the design of the location mapping service can address the last problem. By caching the network address of the addressing bot for each object, the search over the addressing bot overlay network can be minimized. In the same example, suppose there are $k$ objects to be downloaded, the addressing bot overlay has $n_1$ bots, and the region bot overlay has $n_2$ bots. The objects on the first grid will be found with $O(n_2^{1/2})$ hops searching for the first region bot in charge of the grid. Then, the nearby grids will be accessed to find more objects within the range of a region. This will generate around $m$ search hops to communicate with all the neighbor bots. Lastly, with great chance, up to $k$ addressing bots will be contacted for the actual storage location of object resource files. Thus, the total communication overhead for retrieving all the $k$ objects is $O(n_2^{1/2} + m + k)$. Compared with $O(k \cdot log(n_1))$, if $n_1$ and $n_2$ have the same magnitude, the content retrieval approach with the location mapping service has much less communication overhead than the basic approach.

Note that the last strategy actually trades communication overhead in content retrieval with the overhead in cache maintenance. To validate the correct addressing bot for each object, region bots have to periodically query the object resource lookup service. Yet, since addressing bots are reliable nodes in Virtual Net, cache validation rate is expected to be low. How to determine the optimal rate of cache validation will be studied in our future work.

Moreover, although soft bots are constructed by reliable nodes, there is still a chance of node failure or connection failure, which may cause temporary network partition (i.e., the Chord or CAN overlay network). Unexpected overlay network partition may cause requests from one partition unreachable to another partition [Qiu et al., 2007], reducing object availability. Fortunately, existing P2P computing techniques have provided recovery algorithms for handling network partition [Shafaat et al., 2009], [Qiu et al., 2007] by merging partitions once discovered. Thus, the proposed content retrieval strategies are periodically executed. In the case of network partition, unavailable objects could be retrieved in the next cycle as long as the partitions are merged with the underlying recovery algorithm.

---
Algorithm 1. CRC Inventory Construction
---
1. FUNCTION *ConstructCRCInv*(*p*)
2.    $p_r \leftarrow \left(\left\lfloor\frac{p.x}{m\cdot l}\right\rfloor (m\cdot l), \left\lfloor\frac{p.y}{m\cdot l}\right\rfloor (m\cdot l)\right)$
3.    *Regions* ← {$p_r$}
4.    $p_c \leftarrow (p_r.x + (m\cdot l), p_r.y + (m\cdot l))$
5.    IF *p.x* < $p_c.x$ / 2 ∧ $p_r.x \neq 0$
6.      $p_{rn1} \leftarrow (p_r.x - m\cdot l, p_r.y)$
7.      *Regions* ← *Regions* ∪ {$p_{rn1}$}
8.    ELSE IF *p.x* > $p_c.x$ / 2 ∧ $p_r.x \neq x_{max}$
9.      $p_{rn2} \leftarrow (p_r.x + m\cdot l, p_r.y)$
10.     *Regions* ← *Regions* ∪ {$p_{rn2}$}
11.    IF *p.y* < $p_c.y$ / 2 ∧ $p_r.y \neq 0$
12.     $p_{rn3} \leftarrow (p_r.x, p_r.y - m\cdot l)$
13.     *Regions* ← *Regions* ∪ {$p_{rn3}$}
14.    ELSE IF *p.y* > $p_c.y$ / 2 ∧ $p_r.y \neq y_{max}$
15.     $p_{rn4} \leftarrow (p_r.x, p_r.y + m\cdot l)$
16.     *Regions* ← *Regions* ∪ {$p_{rn4}$}
17.    IF |*Regions*| = 3
18.     $p_{rn5} \leftarrow (x, y) : x \in Regions.X \land y \in Regions.X \land (x, y) \notin Regions.X$
19.    FOR EACH $p_{rn} \in$ *Regions*
20.     *inv* ← *QueryCAN*($p_{rn}$)
21.     *CRCInv* ← *CRCInv* ∪ {*inv*}
22.    FOR EACH *object* IN *CRCInv*
23.     IF |*object* − *p*| > $d_r$ / 2
24.     Remove *object* from *CRCInv*
25.    RETRUN *CRCInv*

*Figure 11: Algorithm of CRC inventory construction*

## 5 Algorithm Design

To implement the proximity-based content retrieval strategy, a greedy algorithm is devised. The algorithm is periodically executed. In each cycle, it has two steps: 1) the retrieval of content inventories and 2) the retrieval of contents.

### 5.1 Content Inventories Retrieval

Let $p_0 = (x_0, y_0)$ be the search origin. At the start of a cycle, the content inventories are retrieved by querying the CAN overlay with the region coordinate $p_r$. $p_r$ is calculated by $p_r = \left(\left\lfloor\frac{x_0}{l}\right\rfloor l, \left\lfloor\frac{y_0}{l}\right\rfloor l\right)$. To restrict content retrieval within a region centered at user location, the inventory of three neighbor regions are also retrieved. Two of them, denoted by $r_{n1}$ and $r_{n2}$, are the regions closest to $p_0$. The third region is the common neighbor of $r_{n1}$ and $r_{n2}$. With $p_r$, the coordinates of the neighbor regions can be easily

---

Algorithm 2. Proximity-based Content Retrieval
1. FUNCTION *ContentRetrieval* ()
2.   $p_0 \leftarrow$ *CurrentLocation*()
3.   *CRCInv* $\leftarrow$ *ConstructCRCInv*($p_0$)
4.   $p_{g0} \leftarrow \left( \left\lfloor \frac{p_0 \cdot x}{l} \right\rfloor l, \left\lfloor \frac{p_0 \cdot y}{m \cdot l} \right\rfloor l \right)$
5.   $n_0 \leftarrow$ *QueryCAN*($p_{g0}$)
6.   *Visit* $\leftarrow (n_0, p_{g0})$
7.   WHILE *Visit* $\neq \emptyset$
8.     $(n_i, p_{gi}) \leftarrow$ *Visit.Pop*()
9.     *Visit* $\leftarrow$ *Visit* $\setminus \{(n_i, p_{gi})\}$
10.     $d = |p_{gi} - p_0|$
11.     IF $d < d_r / 2$
12.       (*Neighbors*, *Objects*) $\leftarrow$ *Request*($p_{gi}$)
13.       FOR EACH *object* $\in$ *Objects*
14.         *LoadContent*(*object*)
15.       *Contents* $\leftarrow$ *Contents* $\cup$ *Objects*
16.       FOR EACH $(n_j, p_{gj}) \in$ *Neighbors*
17.         *Visit* $\leftarrow$ *Visit* $\cup \{(n_j, p_{gj})\}$
18.   FOR EACH *obj* IN *CRCInv*
19.     IF *obj* NOT IN *Contents*
20.       *object* $\leftarrow$ *QueryCAN*(*obj.OCoord*)
21.       *LoadContent*(*object*)
22.
23. FUNCTION *LoadContent* (*object*)
24.   *OHash* $\leftarrow$ *Lookup*(*LocalInv*, *object.OID*)
25.   *OHash'* $\leftarrow$ *Lookup*(*CRCInv*, *object.OID*)
26.   IF *OHash* $\neq$ *OHash'*
27.     *SearchAndDownload*(*object.OID*)

*Figure 12: Algorithm of proximity-based content retrieval*

calculated. Figure 10 shows an example of region selection. In this example, the content inventory of Region 1, Region 4, Region 5, and Region 6 are retrieved.

After retrieving the region content inventories, they are merged into one, called the cross-region content (CRC) inventory. For each object $i$ in the CRC inventory, its distance $d$ to $p_0$ is calculated. If $d > d_r$, the object will be removed from the CRC inventory. The entire CRC inventory construction process is described in Figure 11.

## 5.2 Content Retrieval

Firstly, through routing in the region bot overlay network, the region bot in charge of the grid containing $p_0$, denoted by $g_0$, is found first. The location mapping service returns a pair ($n_0$, $p_{g0}$) to the user client where $n_0$ and $p_{g0}$ denote the corresponding region bot and coordinate of the grid respectively. On receipt of ($n_0$, $p_{g0}$), the client requests two types of data from $n_0$: the set of objects $B_0$ located on $g_0$ and the neighbor grids of $g_0$. Meanwhile, the client puts $p_{g0}$ into the local cache $P$. On receiving the request, $n_0$ replies the objects located within $g_0$ and 4 pairs $(n_1, p_{g1})$, $(n_2, p_{g2})$, $(n_3, p_{g3})$, and $(n_4, p_{g4})$ to the client, which refers to the 4 neighbor grids and their managing bots. Note that $n_1 \ldots n_4$ do not necessarily refer to four region bots, as one bot may cover multiple grids.

For each received pair $(n_i, p_{gi})$, the client checks the search stop condition for grid $g_i$. First, if $p_{gi}$ is in $P$, it means the grid has already been explored and one stop condition is met. Secondly, if $|p_{g0} - p_0| > d_r$, another stop condition is met, which indicates the search has reached the search area bound. When either stop condition is met, the content search along the route originated from $g_i$ will be terminated.

For each retrieved object, its integrity will be verified with the object hash in the CRC inventory. After all the routes have terminated in the proximity-based content retrieval process, the client checks with the CRC inventory whether all the objects within the range of a region have been retrieved or updated. If there is any missing object, it will be retrieved individually with the object coordinate in the inventory.

For each object, if it is already in the local cache and its *OHash* in the local inventory is identical to the one in the newly retrieved region inventory, then the object of the latest version has already been downloaded and it can be directly loaded from the local cache. Otherwise, the client will query for the logical computer through the object resource lookup service and download the object file from one of the replica nodes. The complete content retrieval algorithm is described in Figure 12.

## 6 Experiments and Evaluation

In this section, the proposed routing and addressing scheme is evaluated by simulation. The content retrieval algorithm described in Section 5 has been implemented in OMNeT++[2], a module-based network simulator. To avoid hardware impact, all experiments are run in the unit of cycles instead of real time. The default values of the simulation parameters are listed in Table 2. In particular, an experiment can be controlled with or without (addressing or region) bot dynamics, i.e., bot join and bot leave the (CAN or Chord) overlay network. If bot dynamics is applied, a random variable is evaluated for determining node join or leave per 10 cycles. Since, bots are reliable nodes, their dynamics are expected to be low. Thus, without loss generality, both probabilities are configured to be 0.1.

Figure 13 shows the composed simulation GUI with 1 client and 100 objects. The red flag represents the virtual objects distributed on the map. The black solid lines represent the trail of client movement. A random walk algorithm is employed to generate the path for client movement. The dashed circle represents the content discovery range, while the solid circle represents the user perception range. The top two components run the object resource lookup resources (i.e., the Chord overlay) and the location mapping service (i.e., the CAN overlay). The left-hand components provide global parameters and record experiment results to facilitate simulation. We have uploaded the simulation code to GitHub[3] for reader verification.

---

[2] OMNeT++ official website: http://omnetpp.org/
[3] Simulation code: https://github.com/sunniel/VirtualNetContentSharing

| Simulation Parameter | Default Value |
|---|---|
| Number of addressing bots | 100 |
| Number of region bots | 100 |
| Number of regions | 20 |
| Number of grids per region | 9 |
| Node dynamics period | 10 cycles |
| Probability of bot join | 0.1 |
| Probability of bot leave | 0.1 |
| Simulation time length | 1000 cycles |

*Table 2: Simulation parameters and default values*

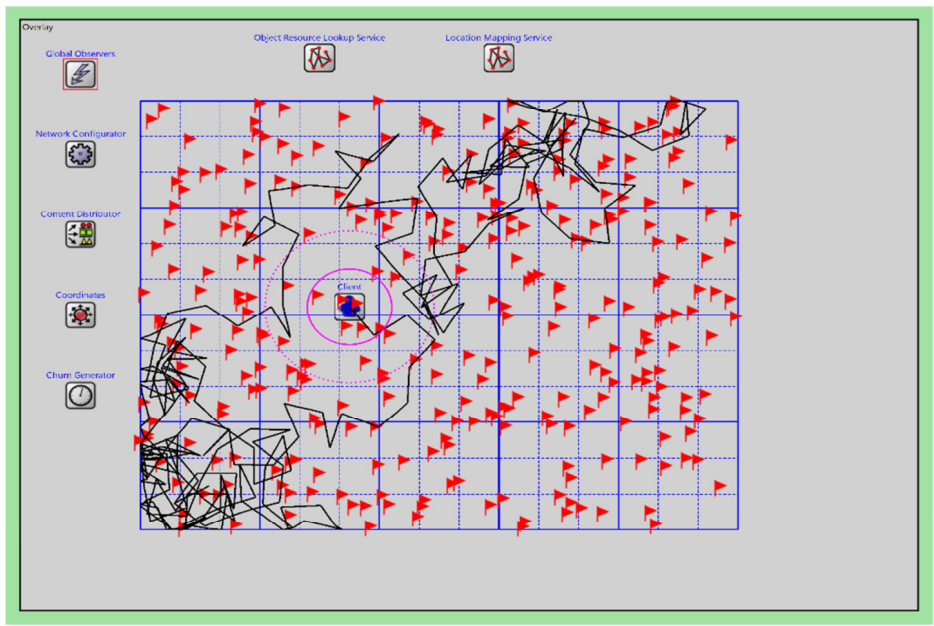

*Figure 13: Virtual world content retrieval simulation with 5×4 regions and 9 grids per region.*

The proposed algorithm is evaluated from communication overhead, content retrieval delay, and load distribution, which are the main achievements of the design. Efficiency and is measured by the number of hops in content retrieval, from retrieving the content inventories to receiving all the objects in one content retrieval cycle. Content retrieval delay is measured by the duration from the start of inventory retrieval to the end of the last object download. Load distribution is measured by the number of routing and forwarding requests on each node within a given length of the period. Moreover, all experiments are tested with and without bot dynamics to study the impact of bot dynamics on system performances.

### 6.1 Communication Overhead

The first experiment evaluates the number of hops per routing as a function of the number of objects. In the experiment, the number of objects in a virtual world is scaled from 100 to 500, and randomly distributed on the map. Though, only the objects located within the range of a region are retrieved, increasing the total number of objects also proportionally changes the number of objects within a region with high probability. The experiment result is shown in Figure 14. Figure 14 (a) describes the content retrieval process without addressing bot dynamics, while Figure 14 (b) describes the process with addressing bot dynamics. It can be found that, the communication overhead increases when addressing bot dynamics is applied, because some cached addressing bots are no longer in charge of the specified objects. Thus, searching over the Chord overlay has to be performed, which increases communication overhead. In both Figures, however, the communication overhead in the improved content retrieval scheme is much lower than that in the basic content retrieval scheme, showing the effectiveness of the proposed strategies. Moreover, the communication cost increases in the improved scheme is also slower than that in the basic content retrieval scheme, meaning that the improved scheme is optimal for virtual world content sharing.

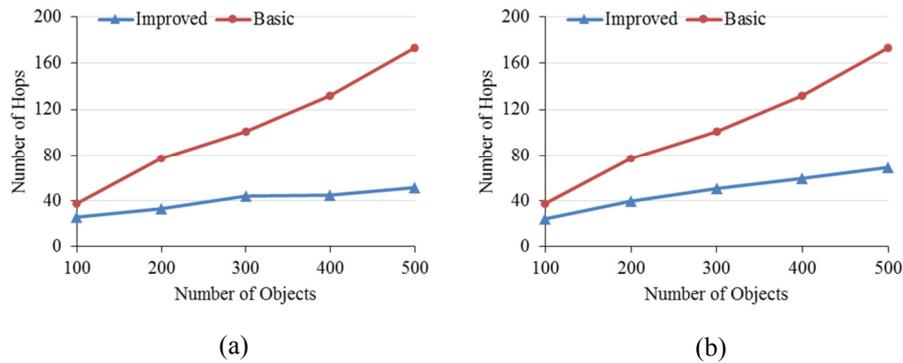

*Figure 14: Number of hops of routing versus number of objects (a) without addressing bot dynamics and (b) with addressing bot dynamics*

### 6.2 Perceived Content Retrieval Delay

The second experiment evaluates the user-perceived delay of content retrieve. The experiment contains three approaches. The improved content retrieval approach follows the sequence of proximate-based content discovery to download the objects. The basic approach follows the sequence of contents listed in the cross-region content inventory. Moreover, the third approach augments the basic approach by sorting the objects in distance to the client position. Figure 15 (a) and (b) shows the experiment result. It is obvious that users can perceive shorter content retrieval delay in the proximate-based approaches than in the inventory-based approach (i.e., the basic approach), because, in the former class, more objects within the user perception range can be retrieved and loaded to the screen first, reducing the time of waiting in play. In this experiment, the proposed approach shows the similar performance to the third approach. To this point, it seems that the basic design with distance sort is simpler than the proposed model. However, the last experiment shows that the basic model will generate more communication overhead than the proposed model. Thus, combining the two results, the proposed model is optimal.

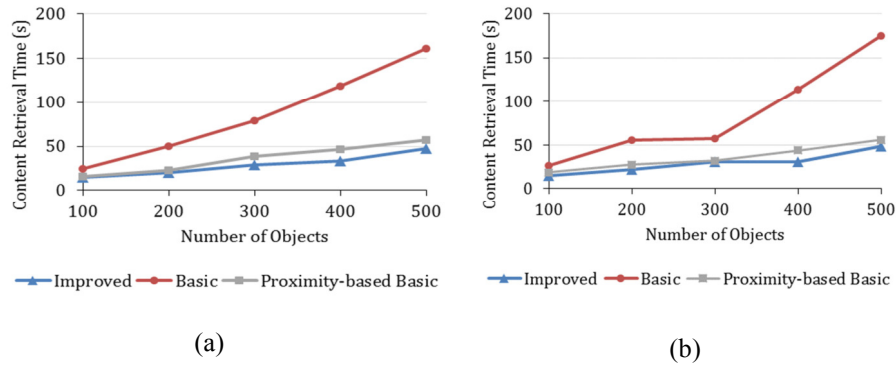

*Figure 15: Content retrieval time versus number of objects (a) without addressing bot dynamics and (b) with addressing bot dynamics*

### 6.3 Load Distribution

The third experiment increases the experiment time 100,000 cycles to study the load distribution on region bots as a function of time. The load on each region bot is quantified and measured by the number of routing request handling and forwarding during a certain period. In the experiment, region bot load is sampled in every 1000 cycles so that 100 samples can be collected in total. Figure 16 shows the trend of the mean load per addressing bot without and with addressing bot dynamics respectively. The results show that both the mean value and the greatest value are large at the beginning of the simulation. With the increase of cycles, load starts decreasing. This is because, with the increase of routing process, more objects are locally cached, unchanged, and do not need to be downloaded again. This result validates the effectiveness of the second strategy that utilizing local cache and content inventory can reduce the number of content retrieval. Moreover, Figure 16 (a) and Figure 16(b) have the similar trends, showing that region bot dynamics does not have large impact on the performance.

## 7 Related Work

The techniques and studies in virtual world content retrieval, P2P file sharing, and virtual world content storage are survey in this section. They are closely related to the work in this paper.

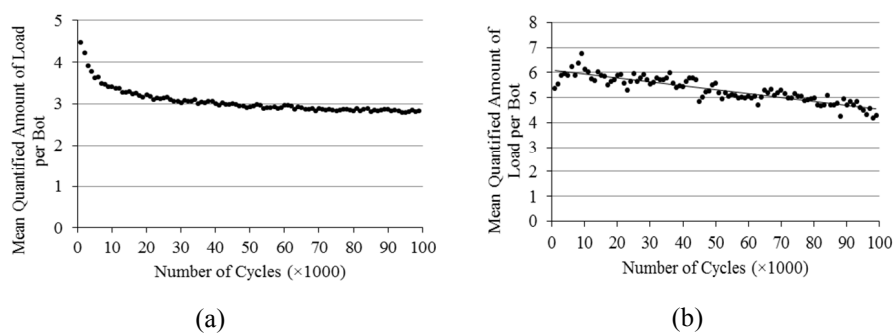

*Figure 16: Region bot load versus number of cycles (a) without addressing bot dynamics and (b) with addressing bot dynamics*

### 7.1   Virtual World Content Retrieval

Existing studies in virtual world content retrieval focuses on reducing the load of content servers in transmitting user-created contents. Compared to MMOGs service, virtual world services needs higher bandwidth for uploading new contents to clients [Symborski, 2008]. To address the issue, [Symborski, 2008] proposed two strategies: local object caching and visibility-based object download restriction. [Symborski, 2010] uses the content reconciliation technique to identify the freshness of local cached objects with summary descriptor. These studies inspire some of the efficient content retrieval strategies in this paper. However, they targeted centralized virtual worlds which have fundamental differences from P2P virtual world. For example, content lists and resource files can be easily retrieved from a server. [Santos et al., 2011] proposes a P2P texture distribution approach by querying the resource files from other user clients in the same region. Thus, the efficiency of the approach heavily relies on the user distribution in a virtual world. Users in the desolate places have to resort to the server for content download. Compared to this approach, our design do not have such concern.

### 7.2   Peer-to-peer File Search

Numerous P2P file search systems [Lua et al., 2005] have been designed and implemented to allow users share files to each other. Most of them focus on the efficiency improvement in searching a shared file. In contrast, virtual world content sharing aims at searching multiple objects within a specific area. The most related work is the P2P semantic file search based on keywords [Gupta et al., 2003] or semantic dimensions [Tang et al., 2003][Haghani et al., 2009]. Different from virtual world content sharing, semantic file search requires many (usually over 100) dimensions to describe a file. Moreover, the semantic file search problem does not require search completeness. Thus, their solutions cannot be applied in virtual world content sharing.

### 7.3   Peer-to-peer Virtual Object Storage

In the early design [Frécon and Stenius, 1998], virtual objects are only maintained by the nodes accessing the objects. If these nodes fail, the object could be lost. To make virtual world persistency, recent approaches [Engelbrecht and Gilmore, 2017], [Holzapfel et al., 2011], [Varvello et al., 2009] store objects on a storage overlay which replicates an object to a fixed number of nodes independent of object access. The storage overlay also handles node failure by creating new replicas of the objects. However, the content of a user is distributed on different nodes, which does not support distributed program execution. To run a virtual world program, the content of a user needs to be retrieved from the storage nodes to the client first. This not only increase communication overhead, the reliability of program execution also relies on the client. If the client fails, user activity will be interrupted. To support both decentralized content storage and program execution, a new redundancy model, called logical computer redundancy (LCR) is proposed in [Shen et al., 2017], which is used as the content storage model in this paper.

## 8   Conclusion

This paper addresses the content sharing issue in virtual world, which is an important function to support collaborative education in a virtual environment. Based on the Virtual Net framework for persistent content storage, we have designed a P2P virtual world content sharing scheme, which allows virtual objects to be efficiently downloaded based on the user location in a virtual world. Our work made the following contribution:

1. A P2P virtual world content sharing and content retrieval scheme has been devised based on the Virtual Net framework.
2. A proximity-based content retrieval strategy has been proposed to reduce the content load delay. The delay could be perceived by users and reduce user experience.
3. Based on the designed content inventory, a local cache verification strategy has been proposed to eliminate the redundancy in content download.
4. A location mapping service has been designed to minimize the communication overhead in storage address query.

The evaluation experiments show the effectiveness of the proposed strategies for improving the efficiency of content sharing. Based on the solution, some prospective and useful applications, such as collaborative mobile immersive education, can be implemented. Our work can be extended to the content sharing problem in augmented reality and mixed reality in which the map is the real-world geography.

The underlying P2P techniques (including Chord, CAN, and Virtual Net) generate additional delay and overheads for reliability [Shen and Guo, 2018], compared with a centralized solution. Yet, we argue that they are the necessary costs to achieve content sharing without a central server. P2P content sharing has its niche and advantages (e.g., scalability and availability) over the centralized counterparty.

Most importantly, it can create a platform without the control from any entity, allowing learning objects on which to be truly persistent and reusable.

The work in this paper is more theoretical than practical in design. We have not implemented the design in a real virtual world application, which may result in some gaps between the design and the actual implementation. We plan to do so in future and run the human subject experiments to further validate the work and check the user experience improvement based on the proposed strategies.

There are some other limitations in the work. First, in the current design, the objects are discovered solely relying on the addressing bots (for both location mapping and content inventory retrieval). This approach can be complemented by peer-assisted object discovery [Ricci et al., 2013] to lighten the load of addressing bots. Peer-assisted object discovery relies on nearby players to discover new objects. We will apply it and seek the balance between these two node discovery mechanisms in the next phase.

Moreover, the objects are assumed to be immobile or rarely moved. But, in a virtual world, some objects are highly dynamic, such as virtual animals and traffic tools. Their positions on a map may change frequently. When an object moves to another region, it has to notify the region bot of the old region and the one of the new region for responsibility transfer. Also, the region-based content inventories maintaining the object have to be updated for content retrieval completeness. Thus, highly dynamic objects will add system and network overhead in location update. To mitigate the issue, dynamic objects could be referred to by their anchor location (e.g., their initial location) instead of their real-time location and they can be discovered through nearby players. This solution will be studied and evaluated in our future work.

**Acknowledgment**

This research is partially supported by the University of Macau Research Grant No. MYRG2017-00091-FST.